\documentclass[prl, twocolumn, showpacs]{revtex4}

\usepackage{graphicx}
\usepackage{amsmath}

\begin{document}

\title{Cavity QED Detection of Interfering Matter Waves}

\author{T. Bourdel, T. Donner,
S. Ritter, A. \"Ottl, M. K{\"o}hl$^\dag$, and T. Esslinger}

\affiliation{Institute of Quantum Electronics, ETH Z\"{u}rich,
H\"{o}nggerberg, CH--8093 Z\"{u}rich, Switzerland}
\date{\today}

\begin{abstract}
We observe the build-up of a matter wave interference pattern from
single atom detection events in a double-slit experiment. The
interference arises from two overlapping atom laser beams
extracted from a Rubidium Bose-Einstein condensate. Our detector
is a high-finesse optical cavity which realizes the quantum
measurement of the presence of an atom and thereby projects
delocalized atoms into a state with zero or one atom in the
resonator. The experiment reveals simultaneously the granular and
the wave nature of matter.
\end{abstract}


\maketitle

The prediction of the duality between particles and waves by de
Broglie \cite{deBroglie} is a cornerstone of quantum mechanics.
Pioneering experiments have observed interferences of massive
particles using electrons \cite{Jonsson1961, Tonomura1989},
neutrons \cite{Zeilinger1988}, atoms \cite{Carnal1991, Keith1991,
Shimizu1992} and even large molecules \cite{Arndt1999}. However,
the simple picture that matter waves show interferences just like
classical waves neglects the granularity of matter. This analogy
is valid only if the detector is classical and integrates the
signal in such a way that the result is a mean particle flux. With
quantum detectors that are sensitive to individual particles the
discreteness of matter has to be considered. The probability to
detect a particle is proportional to the square amplitude of the
wave function and interferences are visible only after the signal
is averaged over many particles.

In the regime of atom optics, single atom detection has been
achieved for example by fluorescence \cite{Hu1994}, using a
micro-channel plate detector for metastable atoms \cite{Wiza1979},
and high-finesse optical cavities \cite{Mabuchi1996}. In these
ex\-per\-iments the size of the de Broglie wave packet of the
particles was much smaller than the detector area. Therefore
localization effects during the detection have been negligible.
With the realization of Bose-Einstein condensation in dilute gases
particles with a wave-function of macroscopic dimensions have
become experimentally accessible. Only very recently the single
atom detection capability has been achieved together with quantum
degenerate samples reaching the regime of quantum atom optics
\cite{Ottl2005,Schellekens2005}. The quantum nature of the
measurement opens perspectives for atom interferometry at and
below the standard limit \cite{Bouyer1997}.

For atoms with a spatially extended wave function, such as in a
Bose-Einstein condensate or in an atom laser beam, a measurement
projects the delocalized atom into a state localized at the
detector \cite{Herkommer1996}. This quantum measurement requires
dissipation in the detection process. For single atom detection we
employ a high-finesse optical cavity in the strong coupling regime
of cavity quantum electrodynamics (QED) \cite{Hood1998,
Munstermann1999, Sauer2004}. We study this open quantum system
including the two sources of dissipation, cavity losses and
spontaneous emission. In particular, we calculate the time needed
for the localisation of an atom in the cavity measurement process.
We then experimentally investigate atomic interferences on the
single atom level using our detector.

A schematic of our experimental setup is shown in figure 1. We
output couple two weak atom laser beams from a Bose-Einstein
condensate and their wave functions overlap and interfere
\cite{Bloch2000}. The flux is adjusted in such a way that there is
on average only one atom at a time in the interferometer. Using
the high-finesse cavity we measure single atom arrival times in
the overlapping beams. We observe the gradual appearance of a
matter wave interference pattern as more and more detection events
are accumulated.

\begin{figure}[htbp]
  \includegraphics[width=0.7\columnwidth, height=0.75\columnwidth]{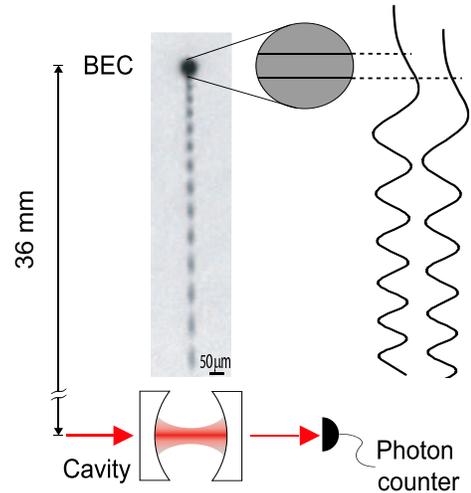}
  \caption{Schematic of the experimental setup. From two well defined regions
  in a Bose-Einstein condensate (BEC), we
  couple atoms to an untrapped state. The real parts of the resulting atom laser
  wave-functions are sketched on the right hand side. The absorption image
  shows an interference pattern corresponding to $\Delta f=1$\,kHz
  and to a flux $\sim 10^6$ times larger than in the actual single atom interference experiment. Monitoring the
  transmission through a high-finesse optical cavity with a photon counter,
  single atom transits are detected.}
 \end{figure}

\begin{figure}[htbp]
  \raisebox{0.06\columnwidth}{\includegraphics[width=0.415\columnwidth]{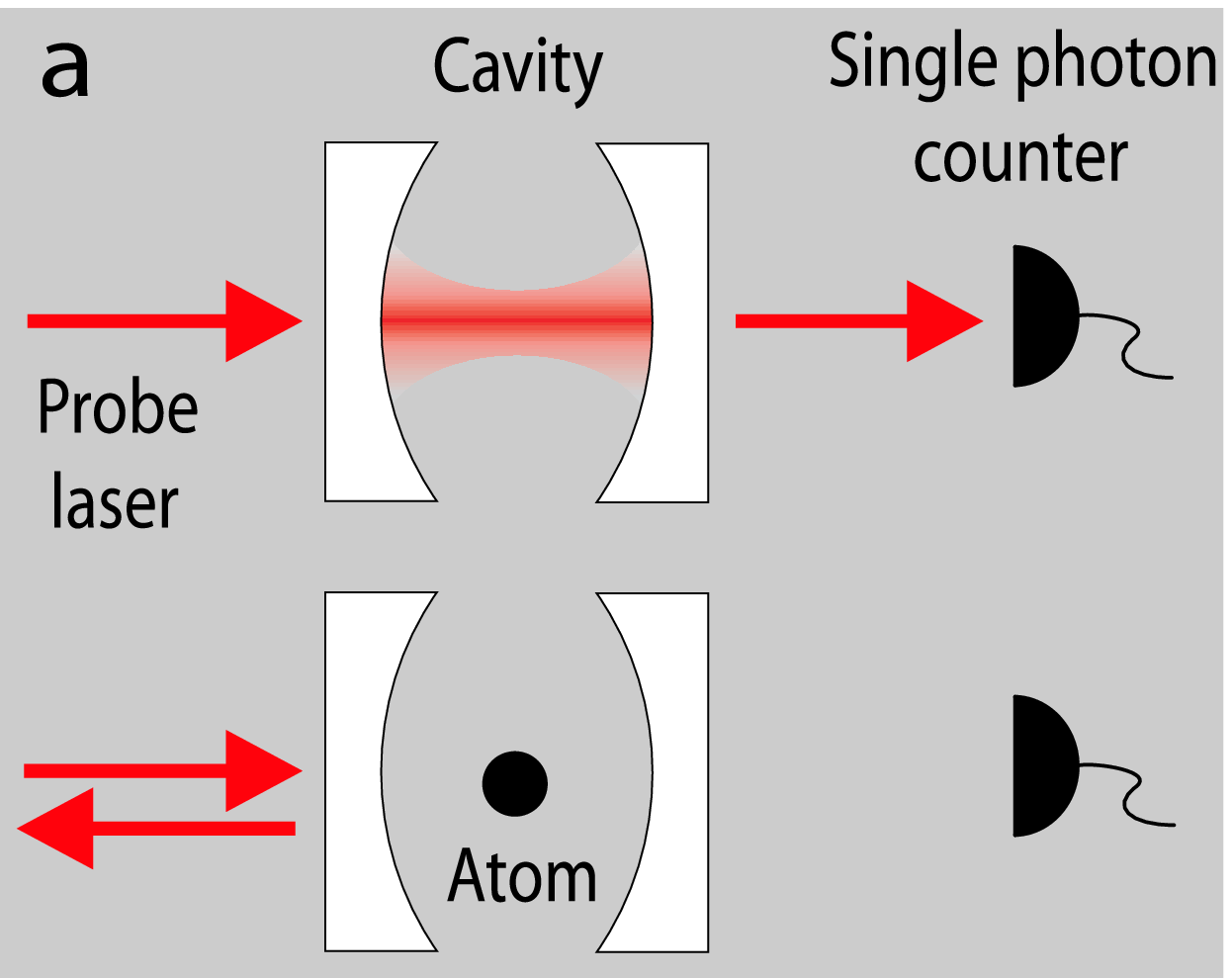}}
  \includegraphics[width=0.565\columnwidth]{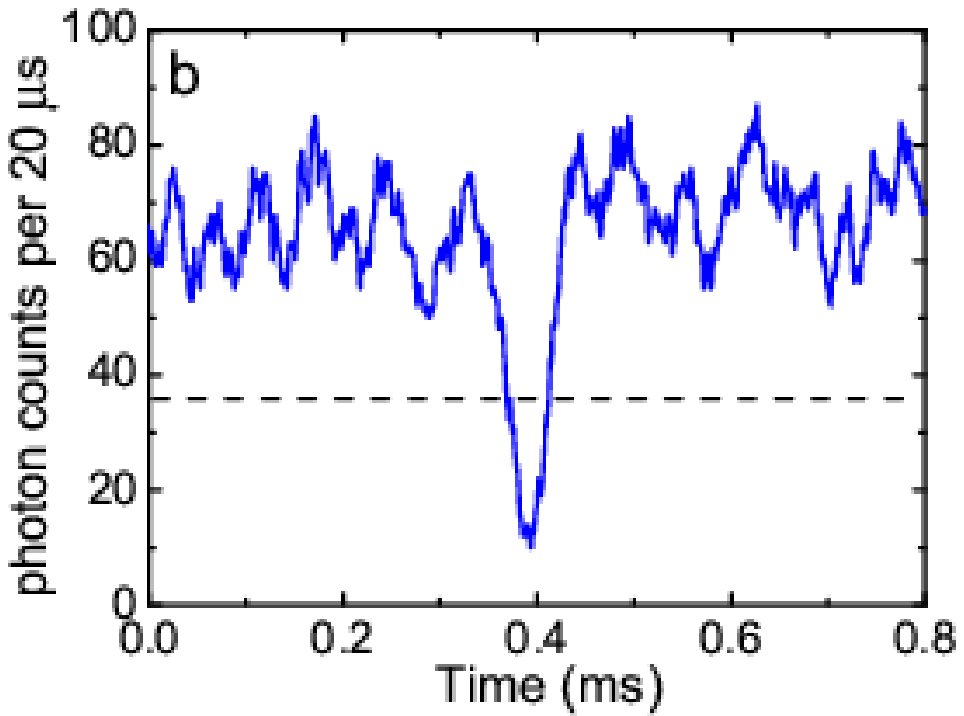}

  \caption{a: Cavity single atom detection principle. An atom detunes
  the high-finesse cavity from resonance and the cavity transmission consequently drops.
  b: Photon flux through the high-finesse optical cavity when an atom
  is detected. The photon count rate is averaged over 20\,$\mu$s.
  The detection threshold is set to be 4 times the
  standard deviation of the photon shot noise (dashed line).}
  \label{fig1}
\end{figure}

Single atom detection in an optical cavity can be captured in a
classical picture: an atom changes the index of refraction in the
cavity and thereby shifts it out of resonance from the probe laser
frequency. In the absence of an atom, the probe beam is resonant
with the cavity and its transmission is maximal. Experimentally we
use a probe power corresponding to five photons on average in the
cavity. The cavity lock is sufficiently stable for the cavity
transmission to be at the photon shot noise limit. The presence of
an atom results in a drop of the cavity transmission (see figure
2). We set the threshold for an atom detection event to a drop in
transmission of four times the standard deviation of the photon
shot noise in our 20\,$\mu$s integration time. Then the overall
detection efficiency of atoms extracted from a Bose-Einstein
condensate is measured to be 0.23(8).

Our cavity has been described in a previous paper \cite{Ottl2005}
and we only recall here its main figures of merit. Its length is
178\,$\mu$m, the mode waist radius is $26\,\mu$m, and its finesse
is $3\times10^5$. The maximum coupling strength between a single
$^{87}$Rb atom and the cavity field $g=2 \pi \times 10.4$\,MHz is
larger than the cavity field decay rate $\kappa= 2 \pi \times
1.4$\,MHz and the atom spontaneous emission rate $\Gamma= 2 \pi
\times 6$\,MHz.  The probe laser and the cavity are red-detuned as
compared to the atomic resonance such that the light force pulls
the atoms to regions where the coupling is large, therefore
enhancing the detection efficiency.

To understand the actual detection process we study the dynamics
of the atom-cavity quantum system taking into account dissipation.
We first consider a classical atom entering a simplified square
shaped cavity mode so that its coupling to the cavity field
increases suddenly to a constant value $g$. The cavity field is
initially coherent with a few photons. We use a two level
approximation for the atom description and assume a 30\,MHz
red-detuning of the probe laser compared to the atomic resonance
\cite{detuning}. In the case of strong coupling the following
dynamics occur. On a short time scale given by $1/g$, the
atom-cavity system exhibits coherent oscillations. It
progressively reaches an equilibrium state on a time scale given
by $1/\kappa$ and $1/\Gamma$ due to cavity loss and atomic
spontaneous emission. These are the two sources of dissipation. In
the equilibrium state, the mean photon number in the cavity is
reduced and the cavity transmission drops.

To evaluate this drop quantitatively, we find the steady-state of
the master equation for the density matrix numerically
\cite{Gardiner1991,Dalibard1992, Molmer1993,Carmichael1993}. For
our parameters, the transmission as a function of the coupling
strength $g$ is plotted in figure 3a. For a maximally coupled atom
$g=2\pi \times 10.4\,$MHz, the average intra-cavity photon number
is found to be reduced from 5 to 0.9, and the number of detected
photons is then reduced by the same ratio. Such a reduction
corresponds well to the largest observed transmission drops. An
example is shown in figure 2. The detection threshold corresponds
to a coupling of $g=2 \pi \times 6.5\,$MHz. Experimentally, unlike
in our model, an atom feels a position dependent coupling as it
transverses the mode profile. However the atom transit time
through the cavity mode ($40\,\mu$s) is long compared to the
cavity relaxation time scales $1/\kappa$ and $1/\Gamma$ and the
atom-cavity system adiabatically follows a quasi equilibrium
state. Therefore the experimental transmission drops can be
compared to the calculated ones.

Specific to our experiment is that a spatially extended matter
wave and not a classical atom enters the cavity. Our system allows
us to realize a quantum measurement of the presence of an atom.
For our low atom flux, we can neglect the probability of having
more than one atom at a time in the cavity. The incoming
continuous wave-function is thus projected into a state with one
or zero atom in the cavity. This projection necessarily involves
decoherence that is introduced by spontaneous scattering and
cavity photon loss. The origin of the decoherence can be
understood as unread measurements in the environment
\cite{Dalibard1992, Herkommer1996}. For example, if a
spontaneously emitted photon is detected, there is necessarily an
atom in the cavity and the wave-function is immediately projected.
Similarly, the more different the light field with an atom in the
cavity is from the field of an empty cavity, the more different is
the scattered radiation out of the cavity, and the projection
occurs correspondingly faster.

\begin{figure}[htbp]
 \includegraphics[width=0.49\columnwidth,clip=true]{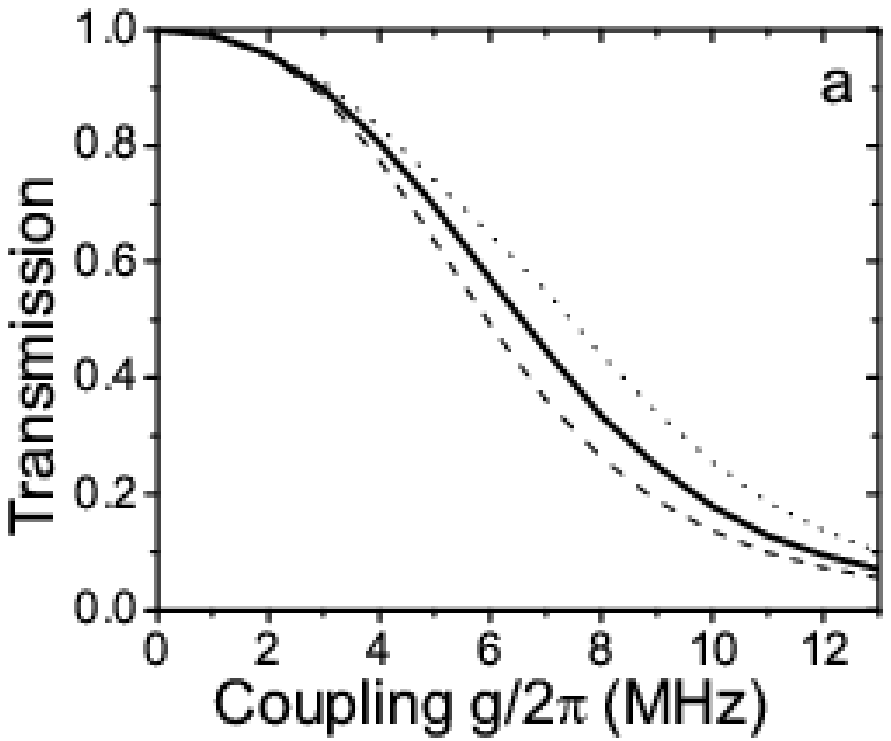}
  \includegraphics[width=0.49\columnwidth,clip=true]{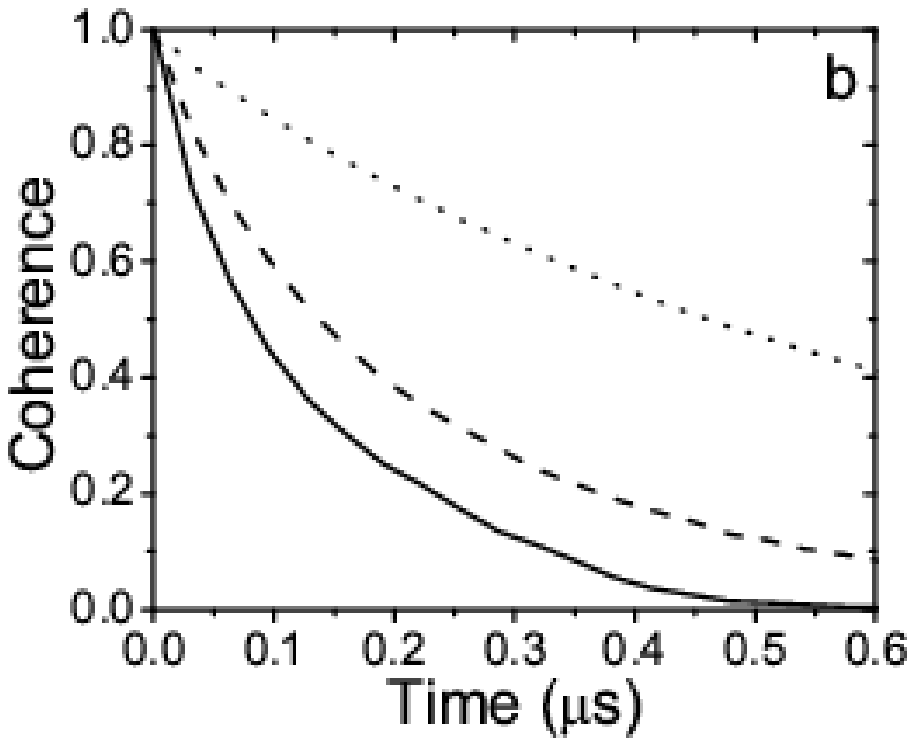}
  \caption{a: Normalized transmission as a function of coupling strength.
  The solid line corresponds to our probe strength of 5 photons
  in the cavity in the absence of an atom. The dashed line is the weak
probe limit. The dotted line corresponds to
  10 photons in the cavity. b: Coherence between the states
  with one and no atom as a function of
  time. The initial coherence is normalized to 1. Solid line:
  $g=2\pi\times 10\,$MHz. Dashed line: $g=2\pi\times 6.5\,$MHz.
  Dotted line: $g=2\pi\times 3\,$MHz.}
\end{figure}

We now quantify the time needed for the projection to occur. For
simplicity, rather than a continuous wave function, we consider a
coherent mixture of one and zero atom entering a square shaped
cavity at a given time. We take the limit when the probability to
have one atom is low. The initial cavity field is the one of an
empty cavity. Dissipation effects are studied by computing the
time evolution of the density matrix \cite{Molmer1993}. The degree
of projection of the initial state can be extracted from the
off-diagonal terms between states with one atom and no atom in the
density matrix. More precisely, we define the coherence as the
square-root of the sum of the squared modulus of the off-diagonal
terms mentioned above. This quantity is maximal for a pure quantum
state with equal probability to have an atom or not. The coherence
is zero for a statistical mixture.

In figure 3b, the temporal evolution of the coherence is plotted.
As expected, it decays to zero at long times due to dissipation.
The decay time increases as the coupling to the cavity is
weakened. In the limit where the cou\-pling vanishes, the
coherence is preserved. The atomic wave function then evolves as
if there was no cavity. For $g>2\pi \times 6.5\,$MHz, the
decoherence time is found to be a fraction of a microsecond. This
value is much lower than the 40\,$\mu$s transit time of an atom
through the cavity and for all our detected atomic transits, the
wave function is thus well projected to a state with one atom. Our
detection scheme realizes a quantum measurement of the presence of
an atom in the cavity. However during an atom transit some photons
are spontaneously scattered and the velocity of the atom is
slightly modified.

Using our cavity detector, we can observe matter wave
interferences on the single atom level. The starting point of the
experiment is a quasi pure Bose-Einstein condensate with
$1.5\times 10^6$ Rubidium atoms in the hyperfine ground state
$|F=1, m_F=-1\rangle$ \cite{Ottl2005}. The atoms are magnetically
trapped with frequencies $\omega_\|=2 \pi \times 7$\,Hz axially
and $\omega_\perp= 2\pi \times 29$\,Hz radially. A weak and
continuous microwave field locally spin-flips atoms from the
Bose-Einstein condensate into the untrapped $|F=2, m_F=0\rangle$
state. This process is resonant for a section of the condensate
where the magnetic field is constant. Because the magnetic moment
of the spin flipped atoms vanishes they fall due to gravity and
form a continuous atom laser \cite{Bloch1999}.

When we apply two microwave fields of different frequencies, we
are able to output couple atom laser beams from two well defined
regions of the condensate \cite{Bloch2000}. The two distinct atom
laser wave functions overlap and interfere. At the entrance of the
cavity, the atomic wave function $\psi$ is well described by the
sum of two plane waves with the following time dependence
\begin{eqnarray}
\psi(t)& \propto & \exp(i \omega_1t)+\exp(i \omega_2t+\phi) \\
       & \propto & \cos((\omega_2-\omega_1)t/2+\phi) \nonumber
\end{eqnarray}
where $\hbar \omega_1$ and $\hbar \omega_2$ are the energies of
the two laser beams and $\phi$ is a fixed phase difference. The
radial dependence of the wave function is neglected. The
probability to detect an atom is given by the square norm of the
wave function which is modulated in time and behaves like a cosine
squared. The modulation frequency of the interference signal is
given by the energy difference between the two atom lasers.
Experimentally, it is determined by the frequency difference of
the two microwave fields and is chosen to be $\Delta f=$10\,Hz,
which corresponds to a distance of 5\,nm between the two output
coupling regions. The two microwave fields are generated such that
the interference pattern is phase stable from one experimental run
to the other.

\begin{figure}[htbp]
  \includegraphics[width=\columnwidth, height=1.3\columnwidth,clip=true]{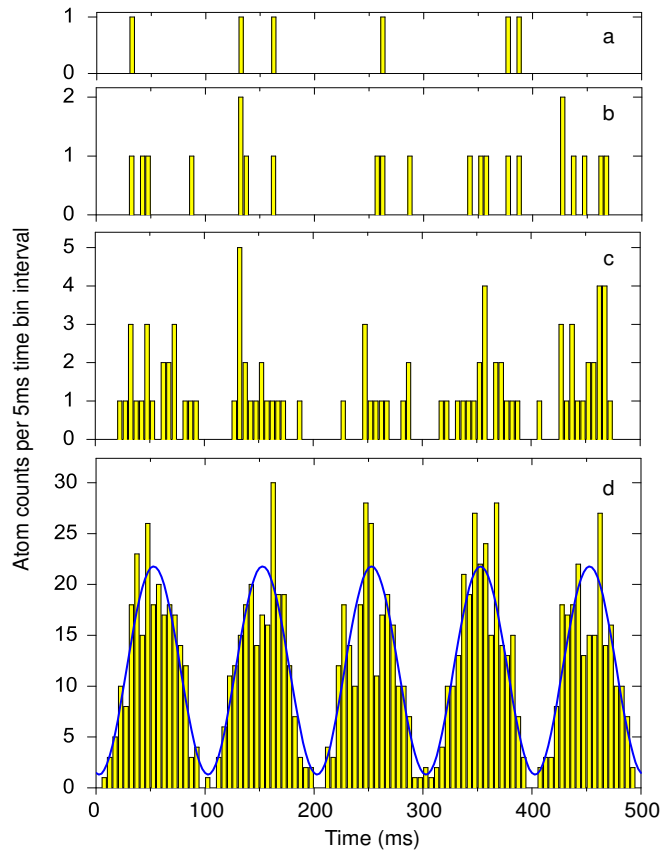}
  \caption{Histograms of the atoms detected in 5 ms time bin intervals. a: Single experimental run.
  b: Sum of 4 runs. c: Sum of 16 runs. d: Sum of 191 runs, the line is a sinusoidal fit. Please note the different scales.}
  \label{fig3}
\end{figure}

The results of the experiment are presented in figure 4. Each
experimental run corresponds to output coupling from a new
condensate. On average $\sim 6$ atoms are detected in 0.5\,s.
After the detection of a few atoms, their arrival times appear to
be random (fig.\,4a). Nevertheless, after adding the results of
several runs, an interference pattern progressively appears
(fig.\,4b-d). The atom number fluctuation is found to be dominated
by the atomic shot noise and the signal to noise ratio of the
interference increases as more data are included. A fit to the
histogram leads to a contrast of 0.89(5). The slight reduction of
contrast is explained by a detected flux of about one atom every 3
runs in the absence of output coupling. We attribute this effect
to artefact detection events and to atoms output coupled from
stray magnetic fields.

We work with a flux of one detected atom per 83\,ms, which is
about the time an atom needs to travel from the condensate region
to the cavity. We are thus in a regime where the atoms fall one by
one in the interferometer. A single atom behaves both like a wave
because its time arrival probability shows an interference pattern
and like a particle as single atoms are detected. This can be
similarly expressed by saying that each individual atom is
released from both slits simultaneously. Our experiment is an
atomic counterpart of the Young's double slit experiment with
individual photons.

To summarize, we detect matter wave interferences with a
high-finesse optical cavity detector which realizes a quantum
measurement of the presence of an atom. We explain how dissipation
plays a crucial role in the detection process and for the
localization of the atom inside the cavity. Using this detector,
we are able to detect a high contrast atom interference pattern at
the single atom level. The coupling of a matter wave to a cavity
QED system opens the route to the quantum control not only of the
internal state of the atoms but also of their positions
\cite{Horak2000}. Using the presented detection technique we can
probe an atomic gas with a good quantum efficiency and introduce
only a minimum perturbation through the measurement. This could
facilitate non-destructive and time-resolved studies of the
coherence of a quantum gas, for example during the formation of a
Bose-Einstein condensate. Similar interference experiments between
two distinct condensates would permit investigations of their
relative phase evolution \cite{Saba2005, Albiez2005} or diffusion
\cite{Javanainen1997}.

We would like to thank G. Puebla for his help with the
simulations. We acknowledge funding by SEP Information Sciences,
OLAQUI (EU FP6-511057), Qudedis (ESF), and QSIT.


\begin{thebibliography}{}
\bibitem{detuning} Our probe beam is linearly polarized orthogonally to the magnetic field direction.
We therefore do not probe on a cycling transition. The 30\,MHz
detuning is an effective value chosen to match the experimental
conditions.
\end{thebibliography}
\end{document}